\documentclass[aps,twocolumn,a4paper,showkeys,footinbib,longbibliography]{revtex4-1}
\usepackage{natbib}
\usepackage{amsmath,amssymb,bm,amsthm}
\usepackage{subfigure}
\usepackage{graphicx}% Include figure files
\usepackage{dcolumn}% Align table columns on decimal point
\usepackage{bm}% bold math
\usepackage[mathlines]{lineno}% Enable numbering of text and display math
\usepackage{color}
\newcounter{one}
\usepackage{url}

\usepackage{textcase}

\usepackage{colortbl}
\usepackage{tabularx}
\usepackage{verbatim}
\usepackage{multirow}

\usepackage[T1]{fontenc} 
\usepackage{lmodern}
\usepackage{bbm}
\usepackage[utf8]{inputenc}
\usepackage{amsfonts}
\usepackage{array}

\usepackage{bbm}
\usepackage{enumitem}
\usepackage{umoline}
\usepackage[usenames,svgnames]{xcolor}
\usepackage{natbib}
\usepackage[hyperindex,breaklinks]{hyperref}
\hypersetup{
     colorlinks=true,       		% false: boxed links; true: colored links
     linkcolor=Navy,          	% color of internal links
     citecolor=Navy,            % color of links to bibliography
     filecolor=Navy,      		% color of file links
     urlcolor=Navy,           	% color of external links
    runcolor=cyan,
 }
\usepackage{graphicx}
\usepackage{amsfonts}
\usepackage{amssymb}
\usepackage{amsmath}
\usepackage{bbm}
\usepackage{enumitem}

\newcommand{\tr}[0]{ {\rm tr}}

\newcommand{\half}[1]{{ \rm h}}
\newcommand{\Oorderof}{\mathcal{O}}
\newcommand{\orderof}[1]{\Oorderof(#1)} 

\newcommand{\for}[0]{\quad \textrm{for} \quad}

\newcommand{\co}{{\rm c}}

\newcommand{\poly}{{\rm poly}}

\newcommand{\Cor}{{\rm Cor}}

\newcommand{\ad}{{\rm ad}}

\def\beq{\begin{equation}}
\def\eeq{\end{equation}}
\def\nbeq{\begin{equation*}}
\def\neeq{\end{equation*}}
\def\<{\langle}
\def\>{\rangle}

\def\tr{{\rm tr}}

\theoremstyle{definition}
\newtheorem{theorem}{Theorem}
\newtheorem{lemma}{Lemma}
\newtheorem{corol}{Corollary}
\newtheorem{assump}{Assumption}

\newcommand{\sectionprl}[1]{{\par\it #1.---}}

\usepackage{ulem}

\begin{document}
\title{Polynomial-time Classical Simulation for One-dimensional Quantum Gibbs States}

\author{Tomotaka Kuwahara}
%%\email{tomotaka.kuwahara@riken.jp}
%\altaffiliation{Present address: Mathematical Science Team, RIKEN Center for Advanced Intelligence Project (AIP),1-4-1 Nihonbashi, Chuo-ku, Tokyo 103-0027, Japan}
\affiliation{
Mathematical Science Team, RIKEN Center for Advanced Intelligence Project (AIP),1-4-1 Nihonbashi, Chuo-ku, Tokyo 103-0027, Japan
}
\affiliation{Department of Mathematics, Faculty of Science and Technology, Keio University, 3-14-1 Hiyoshi, Kouhoku-ku, Yokohama 223-8522, Japan}
\affiliation{Interdisciplinary Theoretical \& Mathematical Sciences Program (iTHEMS) RIKEN 2-1, Hirosawa, Wako, Saitama 351-0198, Japan}

%\affiliation{
%Department of Physics, Graduate School of Science,
%University of Tokyo, Kashiwa 277-8574, Japan
%}
\author{Keiji Saito}
%%\email{saitoh@rk.phys.keio.ac.jp}
\affiliation{Department of Physics, Keio University, Yokohama 223-8522, Japan}

\begin{abstract}
 This paper discusses a classical simulation to compute the partition function (or free energy) of generic one-dimensional quantum many-body systems. Many numerical methods have previously been developed to approximately solve one-dimensional quantum systems. However, there exists no exact proof that arbitrary one-dimensional quantum Gibbs states can be efficiently solved by a classical computer. 
Therefore, the aim of this paper is to prove this with the clustering properties for arbitrary finite temperatures $\beta^{-1}$. We explicitly show an efficient algorithm that approximates the partition function up to an error $\epsilon$ with a computational cost that scales as $n\cdot \poly(1/\epsilon)$, where the degree of the polynomial depends on $\beta$ as $e^{\orderof{\beta}}$. Extending the analysis to higher dimensions at high temperatures, we obtain a weaker result for the computational cost $n\cdot (1/\epsilon)^{\log^{D-1} (1/\epsilon)}$, where $D$ is the lattice dimension.
%Our analysis explicitly clarifies the speciality of one-dimension in terms of computational complexity.

\end{abstract}
\maketitle

\sectionprl{Introduction}
%Hamiltonian complexityの概要、何を目指すか、全体的に何が分かっているか、何が分かっていないか。
One of the central problems in quantum mechanics is the computation of thermodynamic properties in many-body systems. Because of the exponential growth of the Hilbert space, an exact diagonalization approach is inapplicable, unless the system size is small. Hence, for the sake of practicality, it is necessary to resort to an approximation algorithm such as the density matrix renormalization group (DMRG)~\cite{RevModPhys.77.259,PhysRevLett.69.2863} or the quantum Monte-Carlo method~\cite{RevModPhys.73.33,PhysRevB.43.5950}.

In the rapidly developing field of \textit{Hamiltonian complexity}, the computational complexity of various quantum many-body problems has been intensively discussed by evaluating the runtime and precision of  algorithms. 
After the breakthrough by Kitaev~\cite{ref:KempeKitaevRegev}, numerous problems in quantum physics have been found to be intrinsically intractable~\cite{gharibian2015quantum,0034-4885-75-2-022001,cubitt2016complexity,schuch2009computational}, i.e., classified into the quantum Merlin-Arthur (QMA)-complete class. 
More recently, we are on the new stage towards understanding what kind of problem is efficiently computable by a classical computer. 
In this direction, remarkable progress has been made on the efficient classical simulation of 1D gapped ground states~\cite{landau2015polynomial,Arad2017}.

In Hamiltonian complexity, computing thermal equilibrium properties is one of the most important research targets, along with computing the ground state. For a given many-body Hamiltonian $H$, one computes the thermal state $e^{-\beta H}/Z$, where $\beta $ is the inverse temperature and $Z$ is the partition function, $Z=\tr(e^{-\beta H})$. The partition function leads to the free energy per site $n^{-1}\beta^{-1}\log Z $, where $n$ is the system size. Hence, computing $Z$ is obviously the most crucial step to discuss the thermodynamic properties. However, in order to directly calculate $Z$, the trace operation makes it necessary to sum the $\exp[\orderof{n}]$ terms. Hence, a practical computation must rely on some approximation schemes. So far, there have been many empirically successful algorithms for approximating the partition function, based on the quantum Monte-Carlo method~\cite{suzuki1993quantum, suzuki2012quantum}, DMRG~\cite{PhysRevLett.102.190601,PhysRevB.71.241101,PhysRevLett.93.207205,doi:10.1143/JPSJ.66.2221}, quantum belief propagation~\cite{PhysRevB.76.201102,PhysRevA.77.052318,PhysRevB.81.054106,PhysRevLett.106.080403}, etc. More recently, there have been several attempts to implement efficient sampling using a quantum computer~\cite{PhysRevLett.105.170405, temme2011quantum,Yung754,PhysRevLett.103.220502}. Although these algorithms work well empirically, there has been no guarantee of precision. 

%%%
\begin{figure}[t]
\centering
{
\includegraphics[clip, scale=0.4]{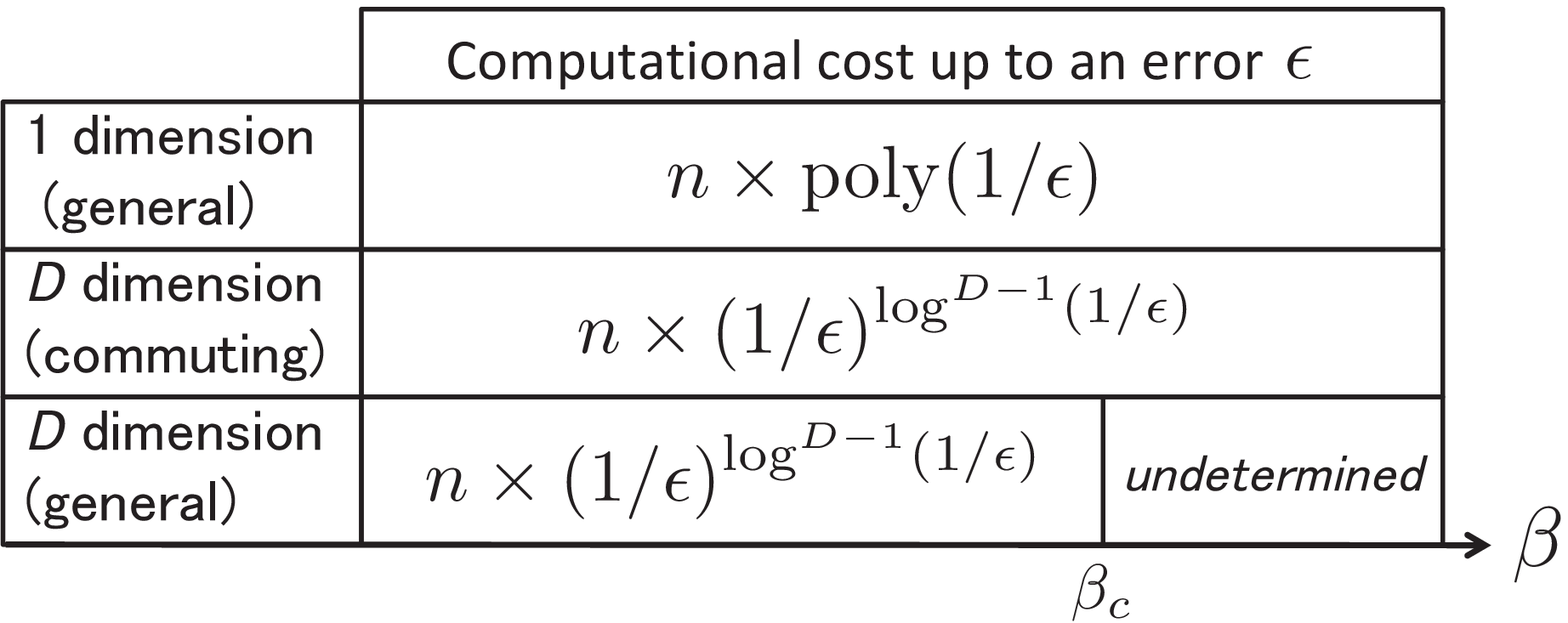}
}
\caption{
  Computational cost to approximate partition function up to error $\epsilon$. 
  In higher dimensions, the theory is valid for high-temperature regime $\beta < \beta_c$ ($=\orderof{1})$.
}
\label{Table_comp}
\end{figure}
%%%

In order to obtain a clear picture of the trade-off relation between the computational cost and precision, there have been several exact studies on an approximation scheme for the partition function, 
including those involving the classical simulation of the quantum ferromagnetic XY model~\cite{bravyi2015monte,PhysRevLett.119.100503} and quantum Gibbs sampling~\cite{brandao2016finite,Kastoryano2016,kato2016quantum}. 
In this paper, we focus on the one-dimensional Gibbs states at an {\it arbitrary temperature}, and provide the first exact evidence of efficient computability in a wide class of one-dimensional quantum systems. We remark here that one-dimensional quantum systems are one of the most actively studied objects, and hence their numerical computation~\cite{PhysRevLett.102.190601,doi:10.1143/JPSJ.66.2221,PhysRevB.76.201102,PhysRevE.89.062137} and computational complexity ~\cite{kato2016quantum,berta2017thermal,barthel2017one,kim2017markovian,PhysRevLett.106.080403} have been intensively discussed. Nevertheless, with the exception of some special cases~\cite{barthel2017one}, the existence of an efficient algorithm that works in polynomial time has not yet been explicitly provided. Here, by the term “efficient algorithm,” we mean an algorithm that approximates $n^{-1}\log Z$ up to an error $\epsilon$ with a computational cost of $\poly(n) \cdot \poly (1/\epsilon)$. In this paper, we show such an efficient algorithm for one-dimensional Gibbs states at an arbitrary finite temperature. 
This is of clear importance because it implies that the decision problem of the free energy in a width of $1/\poly(n)$ is classified into P.

A standard way to tackle this problem is based on the quantum approximate Markov property~\cite{Fawzi2015, kato2016quantum}, which implies that the conditional mutual information between two regions $A$ and $C$ conditioned on the middle region $B$ decays exponentially with a length of $B$. This is an extended version of the clustering properties (i.e., the exponential decay of the bipartite correlation between two separated operators). This property is believed to be true even though the complete proof has not been given; indeed, Kato and Brandao have proved it in a weaker way~\cite{kato2016quantum} for generic one-dimensional Gibbs states. In addition, based on the quantum approximate Markov property, a decision problem of the free energy $n^{-1}\beta^{-1}\log Z$ is proved to be at least in the class NP~\cite{kim2017markovian}. In our analysis, however, we do not rely on the quantum approximate Markov property but employ several elementary techniques, i.e., the standard clustering property~\cite{Araki1969}, quantum belief propagation~\cite{PhysRevB.76.201102}, locality of temperature~\cite{PhysRevX.4.031019} and imaginary time Lieb-Robinson bound.
Moreover, we extend the method to higher dimensional systems in a weaker way than for one-dimensional cases (see Fig.~\ref{Table_comp}).

%
%
%

%We

%マルコフ性を仮定すると、エントロピーの計算などが言える。
%Hastingsの研究とかも
%
%Hamiltonian Complexity~\cite{ref:KempeKitaevRegev,gharibian2015quantum,0034-4885-75-2-022001,cubitt2016complexity}
%
%Contrary to the present , We here show that only the clustering theorem is enough to analyze the computational complexity of thermodynamics values.
%
%W3 prove that many of the interesting quantities such as the partition function can be classically simulated.
%
%by Hastings, where the quantitative analysis of the error was raised as the most important issue 
%
%Classical cases: by using transfer matrix 
%
%Truncation formula, Belief propagation, and imaginary Lieb-Robinson bound
%
%

%
%
\sectionprl{Setup}
We consider a quantum system with $n$ sites defined on a one-dimensional lattice, where each site has a $d$-dimensional Hilbert space.
For instance, spin-$1/2$ systems have $d=2$.  
We consider the following system Hamiltonian which consists of $n$ local terms: 
\begin{align}
H= \sum_{j=1}^n h_j \quad {\rm with} \quad \|h_j \|\le 1 \, .
\label{eq:1D_ham}
\end{align}
Here, each of $\{h_j\}_{j=1}^n$ acts on at most $k$ adjacent sites $\{j,j+1,\ldots, j+k-1\}$ (i.e., up to $k$-body interaction), and $\|\cdots\|$ denotes the operator norm.
(See~\footnote{E.g., consider the spin-$1/2$ chain ($d=2$) described by
 $H=\sum_{j=1}^{n} \left[ J_1 {\bm \sigma}_{j}\cdot {\bm \sigma}_{j+1} +J_2 \sum_{\alpha=x,y,z} \sigma_{j}^{\alpha} {\bm 1}_{j+1} \sigma_{j+2}^{\alpha} \right]$,
  where ${\bm \sigma}_j$ is the Pauli vector matrix at the $j$th site. In this case, one can assign the local Hamiltonian with $k=3$ as
  $h_j =J_1 {\bm \sigma}_{j} \cdot {\bm \sigma}_{j+1}+J_2 \sum_{\alpha=x,y,z} \sigma_{j}^{\alpha} {\bm 1}_{j+1} \sigma_{j+2}^{\alpha}$.
 This framework enables us to write the one-dimensional systems in a unified way, including a case with weak long-range interactions. Without loss of generality, we can choose the parameters $(J_1,J_2)$, so that the operator norm is smaller than $1$.
}
as a specific example of this setup.)

We define the partition function for Hamiltonian $H$ with inverse temperature $\beta$ as follows:
\begin{align}
Z:= \tr (e^{-\beta H}) \, .
\end{align}
The Gibbs state is given by $\rho:=e^{-\beta H}/Z$.
Our main problem is calculating $n^{-1}\log (Z)$ with the desired approximation error $\epsilon$.

For the construction of $Z$, we first define $\{H_i\}_{i=1}^n$ and $\{Z_i\}_{i=1}^n$ as follows:
\begin{align}
H_i:= \sum_{j=1}^{i-1} h_j, \quad Z_i:= \tr \left[ e^{-\beta (H_i + h_i)}\right] \, .
\label{eq:1D_ham_i}
\end{align}
 We then formally decompose the contributions of $H_i$ and $h_i$ in $Z_i$ in the following form:
\begin{align}
 Z_i = \tr \left[ B_ie^{-\beta H_i}B_i^\dagger\right] = \tr \left[ e^{-\beta H_i}A_i \right], \label{replace_by_A_i}
\end{align}
where $A_i :=B_i^\dagger B_i $, which contains information on the operator $h_i$. Eq.~\eqref{replace_by_A_i} separates the contribution of $h_i$ from the Gibbs operator $e^{-\beta (H_i + h_i)}$.
When the local Hamiltonians $\{h_j\}_{j=1}^n$ do not commute with each other, there are several ways to express $A_i$ or $B_i$, such as the Dyson expansion and quantum belief propagation. 
In our analysis, we mainly use a description based on the quantum belief propagation~\cite{PhysRevB.76.201102} (see Eq.~\eqref{eq_QBP} below).

From Eq.~\eqref{replace_by_A_i}, we formally write $Z_i$ as
\begin{align}
 Z_{i} & = Z_{i-1} \tr \left (\rho_i A_{i} \right ) \, , ~~
  \rho_i := e^{-\beta H_{i}}/ Z_{i-1} \, .
\end{align}
This leads to the expression of the partition function in the following form:
\begin{align}
Z = d^{n} \prod_{i=1}^{n} \tr \left (\rho_i A_{i} \right ) \, ,
\end{align}
where we used the relations $Z_n=Z$ and $Z_0=\tr (\hat{1})=d^n$. 
In this framework, our task is reduced to estimating the computation cost to calculate $\tr \left (\rho_i A_{i} \right )$.
Then, the total computation cost is the runtime of computing $\tr \left (\rho_i A_{i} \right )$ multiplied by the system size $n$.

\sectionprl{Assumption and main result}
Throughout this paper, we assume the clustering property of $\rho$ with a fixed $\beta<0$ for arbitrary Hamiltonians $H$. 
More precisely, we assume the following statement:
\begin{assump} \label{clustering property}
 When the Hamiltonian is given in the form of Eq.~\eqref{eq:1D_ham}, the Gibbs state $\rho:=e^{-\beta H}/Z$ ($\beta < \infty$) satisfies 
\beq
\Cor_{\rho} (O', O) \le e^{-l/\xi},\quad \xi < \infty \label{ineq:clustering property}
\eeq
for any two operators $O$ and $O'$ separated by a distance $l$ ($\|O\|=\|O'\|=1$). Here, $\Cor_{\rho}(O',O)$ is the standard operator correlation, that is, 
\beq
\Cor_{\rho}(O',O):= \tr (\rho O'O) - \tr (\rho O')\, \tr (\rho O).
\eeq
\end{assump}
The clustering property implies that a phase transition does not occur at any finite temperature in one-dimensional Gibbs states~\cite{1367-2630-17-8-085007}. This has been rigorously proved by Araki~\cite{Araki1969} for an infinitely large system size $n$. 
The correlation length $\xi$ usually depends on $\beta$ and can be infinite in the limit of $\beta\to \infty$; in one-dimensional Gibbs states, $\xi$ is believed to be at most $e^{\orderof{\beta}}$~\cite{kato2016quantum}.
Under Assumption~\ref{clustering property}, we prove the following statement:
\begin{theorem} \label{main_theorem_contraction}
For a fixed $\beta<\infty$, there exists an approximation scheme of $n^{-1}\log Z$ up to an error $\epsilon$ with the runtime bounded from above by $n\times \poly(1/\epsilon)$. 
\end{theorem}
The power exponent of $\poly(1/\epsilon)$ is at most $\xi e^{\orderof{\beta}}$, which is on the order of $e^{\orderof{\beta}}$ if $\xi < e^{\orderof{\beta}}$.
As the Gibbs state approaches the ground state, the computational cost increases and the partition function may be intractable for $\beta \to \infty$. 
This is consistent with the fact that the computational class to simulate one-dimensional ground state $(\beta \to \infty)$ can be QMA-complete~\cite{Aharonov2009,Bausch2017}.

 \begin{figure}
\centering
{
\includegraphics[clip, scale=0.44]{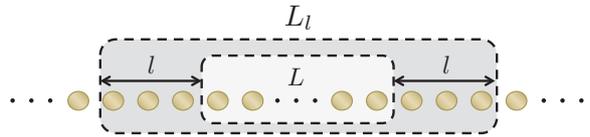}
}
\caption{Schematic for definition of $L_l$ with $l=3$.}
\label{fig:L_l}
\end{figure}

\sectionprl{Approximation of local observables}
The main goal of our classical simulation scheme is to approximate nonlocal operators in $\tr (\rho_i A_i)$ using locally defined operators. 
Hence, we first show how to approximate an expectation value for a locally defined observable. For this purpose, we introduce a local truncation of operators.
 We denote by $L \subset \{1,2,\ldots, n\}$ the subset of adjacent sites that supports the local Hamiltonian $h_i$ (note that $|L|\le k$).
In addition, we define the subset $L_l$, which covers the set $L$ with at most $l$ additional sites at both ends. See Fig.~\ref{fig:L_l} for this situation. Then, for an arbitrary operator $O$, we define $O^{(L_l)}$ as the truncated version of $O$:
\begin{align}
O^{(L_l)}:= \frac{1}{\tr_{L_l^\co} (\hat{1}_{L_l^\co})} \tr_{L_l^\co} (O) \otimes \hat{1}_{L_l^\co} \, , \label{truncation_in_a_region_from_l}
\end{align}
where $L_l^\co$ is the complementary subset to the set $L_l$ (i.e., $L_l^\co:= \{1,2,\ldots, n\}\setminus L_l$). The symbol $\tr_{L_l^\co}$ implies a partial trace with respect to subset $L_l^\co$, and $\hat{1}_{L_l^\co}$ is the identity operator in the Hilbert space for the sites in set $L_l^\co$. Note that $O^{(L_l)}$ is equal to $O$ if the operator $O$ is originally supported in subset $L_l$.

 We consider an arbitrary operator $O_L$ acting on subset $L$ and give an approximation scheme for the expectation value of $\tr ( \rho_i O_L)$. Using the definition~\eqref{truncation_in_a_region_from_l}, we formally decompose the Hamiltonian $H_i$ into $H_i=H_i^{(L_l)} + H_i^{(L_l^\co)} + H_{\partial}$ with $H_{\partial}:=H_i- H_i^{(L_l)} -H_i^{(L_l^\co)} $. Note that the term $H_{\partial}$ is an interaction part between the boundaries of $L_l$ and $L_l^\co$. Hence, in the one-dimensional systems, $\|H_{\partial}\|=\orderof{1}$. We then define $\rho_i^{(L_l)}:= e^{-\beta H_i^{(L_l)}}/\tr (e^{-\beta H_i^{(L_l)}})$ and try to approximate $\tr ( \rho_i O_L)$ by $\tr ( \rho_i^{(L_l)} O_L)$. We find the relation between these quantities using the truncation formula in Ref.~\cite{PhysRevX.4.031019} (see also~\cite{Supplement_1D}):
\begin{align}
\tr ( \rho_i O_L) &= \tr ( \rho_i^{(L_l)} O_L) \notag \\
 &- \beta \int_0^1 \int_0^1 \Cor_{\rho(s)} (O_L \, , H_{\partial} (s, \kappa )
% \rho_i(s)^{-\kappa} H_{\partial } \rho_i(s)^{\kappa}
 ) ds d\kappa, \label{approx_truncation_formulae}
\end{align}
where $\rho_i(s):= e^{-\beta H_i(s)}/\tr (e^{-\beta H_i(s)})$ with $H_i(s):=H_i- sH_{\partial}$, and the operator $H_{\partial}(s,\kappa)$ is defined as
\begin{align}
 H_{\partial}(s,\kappa):= \rho_i(s)^{-\kappa} H_{\partial} \rho_i(s)^{\kappa} \, .
\end{align}
The second line in Eq. \eqref{approx_truncation_formulae} provides the approximation error when using $\tr ( \rho^{(L_l)} O_L)$.

%By using \eqref{approx_truncation_formulae}, we can estimate the approximation error by $\tr ( \rho^{(L_l)} O_L)$.

\sectionprl{Commuting case}
We first consider the simple case where the local Hamiltonians $\{h_j\}_{j=1}^n$ commute with each other, i.e., $[h_i, h_j] = 0$ for $\forall i,j \in \{1,2,\ldots, n\}$.
In this case, as in classical spin chains, we can exactly calculate partition function $Z$ using the transfer matrix technique with a computational cost of $n e^{\orderof{k}}$.
Although this case is very simple, it provides a useful picture to prove Theorem~\ref{main_theorem_contraction} for the general case given in the next section.

From the commutability of local Hamiltonians, operator $A_i$ is given by $A_i = e^{-\beta h_i}$, and $H_{\partial}(s,\kappa)= H_{\partial}$, which drastically simplify the problem.
%is now defined on the subset $L=\{i,i+1,\ldots,i+k-1\}$ from the definition of $h_i$.Then, by using the fact that $\rho(s)^{-\kappa} H_{\partial } \rho(s)^{\kappa} = H_{\partial }$,
Then, from Eq. (\ref{approx_truncation_formulae}) with $O_L=A_i$, we have the relation
$
\tr (\rho_i A_i) = \tr ( \rho_i^{(L_l)} A_i) - \beta \int_0^1 \int_0^1 \Cor_{\rho_i(s)}(A_i,H_{\partial }) ds d\kappa 
$.
Under the condition~\eqref{ineq:clustering property}, we
 note the following relation:
\[\Cor_{\rho_i(s)}(A_i,H_{\partial }) \le e^{\beta \|h_i\|} \|H_\partial\| e^{-l/\xi} \le C e^{\beta} e^{-l/\xi} \]
where $C$ is a constant of $\orderof{1}$.
This immediately leads to
\begin{align} 
\left|\tr (\rho_i A_i) - \tr ( \rho_i^{(L_l)} A_i) \right| \le C \beta e^{\beta} e^{-l/\xi}.\label{approx_local_ob}
\end{align}
Because subset $L_l$ has $\orderof{l}$ sites, the computational cost to calculate $\tr ( \rho_i^{(L_l)} A_i)$ is at most $d^{\orderof{l}}$.
In order for the error $\beta e^{\beta} e^{-l/\xi}$ to be smaller than $\epsilon$ for a fixed $\beta$, we need to take $l=\orderof{\xi \log \epsilon^{-1}}+\orderof{\xi \beta}$. Thus, for a sufficiently small $\epsilon$, this gives the computational cost of $(1/\epsilon)^{\orderof{\xi} \log d}$. 
This completes the proof of Theorem~\ref{main_theorem_contraction}.

\begin{figure}[t]
\centering
{
\includegraphics[clip, scale=0.33]{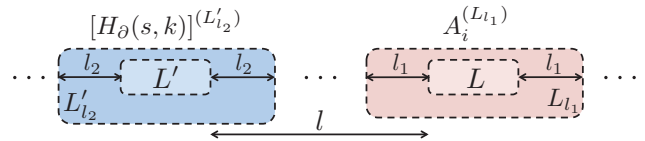}
}
\caption{(color online) Schematic representation of $A_i^{(L_{l_1})}$ and $[H_{\partial}(s,\kappa)]^{(L_{l_2}')}$.}
%\caption{(color online) The approximation of $\tr(\rho_i A_i)$ by $\tr ( \rho_i^{(L_l)} A_i)$ is ensured by the truncation formula~\eqref{truncation_in_a_region_from_l}. However, in order to apply the clustering properties for $ \Cor_{\rho_i(s)}\left[A_i, H_{\partial }(s,\kappa)\right]$, we need to prove that $A_i$ and $H_{\partial }(s,\kappa)$ are approximated by locally truncated $A_i^{(L_{l_1})}$ and $[H_{\partial}(s,\tau)]^{(L'_{l_2})}$. }
\label{fig:truncation}
\end{figure}

\sectionprl{General case}
We next consider the general case where the local Hamiltonians do not commute with each other. The basic strategy to arrive at Theorem 1 is the same as in the commuting case. The main difference from the commuting case is that both of the operators $A_i$ and $H_{\partial}(s,\kappa)$ are no longer local operators. We use the expression of operator $A_i=B_i B_i^\dagger$ based on the quantum belief propagation \cite{PhysRevB.76.201102}, where $B_i$ is written as 
\begin{eqnarray}
% \begin{split}
 B_i &=& \mathcal{T}_\tau \Bigl[e^{\int_0^1 \eta_i(\tau) d\tau } \Bigr] \, , \label{eq_QBP} \\ 
 \eta_i (\tau) &=& \frac{-\beta h_i}{2} - i \sum_{n\ge 1} \int_{-\infty}^{\infty} {\rm sign}(t) e^{-{2\pi n t\over \beta}} h_i(t,\tau)dt \, , ~~~~~
 % \notag 
 %\end{split}
\end{eqnarray}
where $h_i(t,\tau) := e^{i(H_i +\tau h_i) t} h_i e^{-i (H_i +\tau h_i) t}$, and $\mathcal{T}_\tau$ is an ordering operator with respect to $\tau$.

The crucial point to show Theorem \ref{main_theorem_contraction} is that operators $A_i$ and $H_{\partial}(s,\kappa)$ can still be treated as {\it local} even in the general case.
%To see this point, we write $\tr (\rho_i A_i)$ with local variables and estimate the approximation error.
To show this, we define subsets $L'$, which supports the interaction part $H_{\partial}$, and $L_{l_2}'$ with additional $l_2$ sites in the same way as defined for $L_l$ in Fig.~\ref{fig:L_l}. Using the same truncation scheme~\eqref{approx_truncation_formulae}, we define $[H_{\partial}(s,\tau)]^{(L'_{l_2})}$. Then, we show that $A_i$ and $H_{\partial}(s,\kappa)$ can be approximately replaced by $A_i^{(L_{l_1})}$ and $[H_{\partial}(s,\tau)]^{(L'_{l_2})}$, respectively. See the schematic representation of these in Fig.~\ref{fig:truncation}. 
To estimate the approximation error, we start with the following inequality: 
 \begin{align} 
 &\left| \tr(\rho_i A_i) - \tr( \rho_i^{(L_l)} A_i^{(L_{l_1})} )\right| \le \| A_i - A_i^{(L_{l_1})}\| \notag \\
&+ 2\beta \int_0^1 \int_0^1\biggl( \bigl \|A_i^{(L_{l_1})}\bigr\| \cdot \bigl \| H_{\partial}(s,\kappa) - [H_{\partial} (s,\kappa) ]^{(L'_{l_2})} \bigr\| \notag \\
&+\frac{1}{2}\left | \Cor_{\rho_i(s)}\left( A_i^{(L_{l_1})} ,[H_{\partial} (s,\kappa) ]^{(L'_{l_2}) }\right) \right | \biggr) ds d\kappa. \label{approx_A_i_rho_decomp}
\end{align}
To derive this inequality, we insert $O_L=A_i^{(L_{l_1})}$ in Eq.~\eqref{approx_truncation_formulae} and estimate the upper bound of the left hand side of Eq. \eqref{approx_A_i_rho_decomp}.
We then consider the upper bound of each term and estimate the conditions of $(l_1,l_2,l)$ to obtain the desired approximation error $\epsilon$. 
To this end, we consider the following three steps:(i) we compute the bound of $\| A_i - A_i^{(L_{l_1})}\|$, 
(ii) we compute the bound of $\| H_{\partial}(s,\kappa) - [H_{\partial} (s,\kappa) ]^{(L'_{l_2})} \|$, 
and (iii) we consider the bound of the third term in Eq. \eqref{approx_A_i_rho_decomp}. 
In the first two steps, we derive the conditions on $l_1$ and $l_2$ to achieve the error $\epsilon$. 
In step (iii), we use the clustering property (\ref{ineq:clustering property}) by taking a sufficiently large length $l$.

We first consider step (i).
From the Lieb-Robinson bound~\cite{ref:LR-bound72,PhysRevLett.97.050401}, the operator $h_i(t,\tau)$ is locally approximated by $[h_i(t,\tau)]^{L_l}$ for $l=\orderof{t}$. 
After straightforward but lengthy calculations, we obtain
 \begin{align} 
 \| A_i - A_i^{(L_l)}\| \le C_{k,\beta} \exp\left(\beta-\frac{\pi l}{2e k^3\beta}\right), \label{approx_A_i}
\end{align}
 with $\|A_i\| \le e^{\beta}$ and $\|A_i^{(L_l)}\| \le e^{\beta}$, where $C_{k,\beta}$ is a constant of $\orderof{k^2\beta^2}$.
We present the details of the derivation in the supplementary material~\cite{Supplement_1D}.
The inequality~\eqref{approx_A_i} implies that $A_i^{(L_{l_1})}$ is approximated up to an error $\epsilon$ by choosing $l_1=\orderof{\beta^2} + \orderof{\beta} \log \epsilon^{-1}$.
This choice of $l_1$ also ensures that the expectation $\tr( \rho^{(L_l)} A_i^{(L_{l_1})})$ is approximated up to the error $\epsilon$ by 
 \begin{align} 
\tr( \rho^{(L_l)} A_i^{(L_{l_1})}) = \frac{ \tr\left[e^{-\beta (H_i^{(L_l)} + h_i )}\right] }{ \tr \left[e^{-\beta H_i^{(L_l)}}\right]} + \orderof{\epsilon}. \label{cal_A_i_Z}
\end{align}
We notice that the computational cost to calculate the first term is at most $d^{\orderof{l}}$.

 In step (ii), we consider the local approximation of $H_{\partial}(s,\kappa)$.
The operator $H_{\partial}(s,\kappa)$ is regarded as an imaginary time evolution of $H_{\partial}$ by the Hamiltonian $H_i(s)=H_i -s H_{\partial}$; here, 
the length of the time is $\beta \kappa$. For one-dimensional systems, we can prove the following form of the imaginary Lieb-Robinson bound (see~\cite{Supplement_1D} for the proof):
\begin{lemma} \label{Lem:Imaginary_lieb-Robinson}
For an arbitrary one-dimensional Hamiltonian $H$ in the form of Eq.~\eqref{eq:1D_ham}, we have
 \begin{align} 
 \|O_{L'}(i\tau) - [O_{L'}(i\tau)]^{(L_{l_2}')} \| \le \frac{\zeta_{l_2} ^{\lceil {l_2}/k \rceil}}{1-\zeta_{l_2}}, \label{approx_O_L}
\end{align}
where $O_{L'}(i\tau) := e^{\tau H}O_{L'} e^{-\tau H}$ and
$
\zeta_{l_2} := \frac{6e k\tau}{\gamma \log \lceil {l_2}/k \rceil }
$,
with $\gamma\simeq 1.6026$.
Here, $O_{L'}$ is an arbitrary operator supported in subset $L'$.
\end{lemma}
The proof is given by expanding the operators with respect to $\tau$ using the Baker-Campbell-Hausdorff (BCH) expansion. It should be noted that this lemma is valid only for one-dimensional systems.
   
Using this lemma, in order to approximate $H_{\partial}(s,\tau)$ by $[H_{\partial} (s,\kappa) ]^{(L'_{l_2})}$ up to an error $\epsilon$, we need to choose $l_2= e^{\orderof{\beta}} \log \epsilon^{-1}$.

 We finally consider step (iii). The correlation
between $A_i^{(L_{l_1})}$ and $[H_{\partial} (s,\kappa) ]^{(L'_{l_2})}$ is bounded using the clustering inequality~\eqref{ineq:clustering property}.
The distance between these two operators is $l-(l_1+l_2)$, and Ineq.~\eqref{ineq:clustering property} reads as follows:
\[
\left | \Cor_{\rho_i(s)}\left( A_i^{(L_{l_1})} ,[H_{\partial} (s,\kappa) ]^{(L'_{l_2}) }\right) \right |\le e^{e^{\orderof{\beta}}} e^{[ l-(l_1+l_2)]/\xi}. \]
We now choose $l_1=\orderof{\beta^2} + \orderof{\beta} \log \epsilon^{-1}$ and $l_2= e^{\orderof{\beta}} \log \epsilon^{-1}$. Hence, we have to choose $l$ as
 \begin{align} 
l= \xi e^{\orderof{\beta}} \log \epsilon^{-1}
\end{align}
Thus, from Eq.~\eqref{cal_A_i_Z}, the computational cost is at most $d^{\xi e^{\orderof{\beta}} \log \epsilon^{-1}}$ in order to calculate $\tr(\rho A_i)$ up to the error $\epsilon$. 
This completes the proof of Theorem~\ref{main_theorem_contraction}.

\sectionprl{Extensions to higher dimensional systems}
The same analysis can be applied to higher dimensional systems under the assumption of the clustering~\eqref{ineq:clustering property}. This assumption has been rigorously proven at high temperatures~\cite{PhysRevLett.93.126402, PhysRevX.4.031019}. 
In a case where the Hamiltonian is commuting, only the clustering property determines the computational cost to calculate the partition function; we apply the same approximation scheme as in Eq.~\eqref{approx_local_ob}.
The only difference comes from the number of sites in $L_l$. Because subset $L_l$ has $\orderof{l^D}$ sites, the computational cost to calculate $\tr ( \rho_i^{(L_l)} A_i)$ is at most $d^{\orderof{l^D}}$.
Thus, in order to keep the error smaller than $\epsilon$ for a fixed $\beta$, we need to take $l=\orderof{\xi \log \epsilon^{-1}}$. 
This implies a computational cost on the order of $(1/\epsilon)^{\xi^D \log^{D-1}(1/ \epsilon)}$. 

In non-commuting cases, using Ineq.~\eqref{approx_A_i_rho_decomp} makes it possible to estimate the approximation error. 
The first difference is that we cannot use the belief propagation technique but need to apply the Dyson expansion for the local approximation of $A_i$ as follows~\footnote{
We give the derivation of the Dyson expansion in the supplementary material~\cite{Supplement_1D}.  
Actually, based on the Dyson expansion of Eq.~\eqref{def:Dyson_setup}, we can also derive the main theorem~\ref{main_theorem_contraction} for one-dimensional Gibbs states in a rather weaker way; the power exponent of $\poly(1/\epsilon)$ in Theorem~\ref{main_theorem_contraction} becomes much worse as $\exp[e^{\orderof{1/\epsilon}}]$. }:
\begin{align}
&A_i :=e^{\beta H_{i}} e^{-\beta (H_{i} +h_{i})} =\mathcal{T}_\tau \Bigl[e^{-\int_0^\beta e^{\tau H_{i}} h_{i} e^{-\tau H_{i}} d\tau}\Bigr] \, . \label{def:Dyson_setup}
\end{align}
In this case, we rely on Lemma~\ref{Lem:Imaginary_lieb-Robinson} for the local approximation of $A_i$. 
Then, the power exponent of $\poly(1/\epsilon)$ in Theorem~\ref{main_theorem_contraction} becomes much worse as $\exp [e^{\orderof{\beta}}]$.
The second difference, which is more essential, is that Ineq.~\eqref{ineq:norm_multi_commutator} does not hold, and the convergence of the BCH expansion is not ensured above a critical $\beta_c=\orderof{1}$.
Then, if $\beta<\beta_c$, the same computational cost as in the commuting cases, i.e., $(1/\epsilon)^{\xi^D \log^{D-1}(1/ \epsilon)}$, is derived for the approximation up to the error $\epsilon$.
However, for $\beta\ge \beta_c$, the (quasi-)locality of $A_i$ and $H_{\partial}(s,\kappa)$ in Ineq.~\eqref{approx_A_i_rho_decomp} is no longer ensured, and hence it is not possible to determine the computational complexity of the partition function using only the clustering property. 

%1: decayの方法にどう依存するか。べき減衰はダメ？
%2: βに対して、e^{O(\beta)}で定理が破綻するが、よりbetterな解は存在しうるか？

\sectionprl{Summary and remark on one-dimensionality}
In this letter, we have shown the computational cost to calculate the partition function $n^{-1}\log Z$ up to an error $\epsilon$. We summarize our results in Fig.~\ref{Table_comp}.
In one dimension, our analysis exactly shows that the decision problem of the free energy in a width of $1/\poly(n)$ is classified into P. 

Here, we comment on the special point of one-dimensional systems. 
One of the key relations is shown in Lemma \ref{Lem:Imaginary_lieb-Robinson}. 
This lemma was derived based on the expansion $O_{L'}(i\tau)=\sum_{m=0}^\infty (\tau^m/m! )\ad_{H}^{m} (O_{L'})$. 
For generic Hamiltonians with $k$-body interactions, the norm of $\ad_{H}^{m} (O_{L'})$ is upper-bounded by $(Cm)^m$, with $C$ a constant~\cite{KUWAHARA201696,PhysRevLett.116.120401} (see Lemma~3 in Ref.~\cite{KUWAHARA201696}). 
This scaling implies that above a critical $\tau_c$, the BCH expansion diverges and a local approximation of $O_{L'}(i\tau)$ no longer holds, as mentioned in Ref.~\cite{huang2017lieb}. 
On the other hand, for one-dimensional Hamiltonians, we can prove an improved upper bound for the norm $\| \ad_{H}^{m} (O_{L'})\|$ as follows:
 \begin{align} 
\left\| \ad_{H}^{m} (O_{L'}) \right\| \le \left(\frac{6km}{\gamma \log m}\right)^m. \label{ineq:norm_multi_commutator}
\end{align}
Because of the logarithmic correction $\log m$, the summand $(\tau^m/m!)\ad_{H}^{m} (O_{L'})$ always converges for $m\gtrsim e^{\orderof{k\tau }}$. This point plays a crucial role in the existence of a polynomial algorithm for calculating the quantum partition function.

\begin{acknowledgments}
%T.I. was partially supported by the Program for Leading Graduate Schools, MEXT, Japan. 
The work of T. K. was supported by the RIKEN Center for AIP and JSPS KAKENHI Grant No. 18K13475.
K.S. was supported by JSPS Grants-in-Aid for Scientific Research (JP16H02211).

%T.K. was partially supported by the Program for World Premier International Research Center Initiative (WPI), MEXT, Japan. 
%T.M.'s research was financially supported by JSPS KAKENHI Grant No. 15K17718. 
%N.H.'s research was partially supported by Kakenhi Grants Nos. 15K05200, 15K05207, and 26400409 from Japan Society for the Promotion of Science. 
\end{acknowledgments}

\bibliography{Thermal_1D}

\appendix

\begin{widetext}

%%%%%%%%%%%%%%%%%%%%%%%%%%%%%%%%%%%%%%%%%%%%%%%%%%%%%%%%%%%%%%%%%%%%%%%%%%%%%%%%%%%%%%%%%%%%%%%%%%%%%%%%%%%%%%%%%%%%%%%%%%%%%%%%%%%%%%%%%%%%%%%%%%%%%%%%%%%%%%%%%%%%%%%%%%%%%%%%%%%%%%%%%%%%%%%%%%%%%%%%%%%%%%%%%%%%%%%%%%%%%%%%%%%%%%%%%%%%%%%%%%%%%%%%%%%%%%%%%%%%%%%%%%%%%%%%%%%%%%%%%%%%%%%%%%%%%%%%%%%%%%%%%%%%%%%%%%%%%%%%%%%%%%%%%%%%%%%%%%%%%%%%%%%%%%%%%%%%%%%%%%%%%%%%%%%%%%%%%%%%%%%%%%%%%%%%%%%%%%%%%%%%%%%%%%%%%%%%%%%%%%%%%%%%%%%%%%%%%%%%%%%%%%%%%%%%%%%%%%%%%%%%%%%%%%%%%%%%%%%%%%%%%%%%%%%%%%%%%%%%%%%%%%%%%%%%%%%%%%%%%%%%%%%%%%%%%%%%%%%%%%%%%%%%%%%%%%%%%%%%%%%%%%%%%%
%%%%%%%%%%%%%%%%%%%%%%%%%%%%%%%%%%%%%%%%%%%%%%%%%%%%%%%%%%%%%%%%%%%%%%%%%%%%%%%%%%%%%%%%%%%%%%%%%%%%%%%%%%%%%%%%%%

%Dyson Expansion

%%%%%%%%%%%%%%%%%%%%%%%%%%%%%%%%%%%%%%%%%%%%%%%%%%%%%%%%%%%%%%%%%%%%%%%%%%%%%%%%%%%%%%%%%%%%%%%%%%%%%%%%%%%%%%%%%%%%%%%%%%%%%%%%%%%%%%%%%%%%%%%%%%%%%%%%%%%%%%%%%%%%%%%%%%%%%%%%%%%%%%%%%%%%%%%%%%%%%%%%%%%%%%%%%%%%%%%%%%%%%%%%%%%%%%%%%%%%%%%%%%%%%%%%%%%%%%%%%%%%%%%%%%%%%%%%%%%%%%%%%%%%%%%%%%%%%%%%%%%%%%%%%%%%%%%%%%%%%%%%%%%%%%%%%%%%%%%%%%%%%%%%%%%%%%%%%%%%%%%%%%%%%%%%%%%%%%%%%%%%%%%%%%%%%%%%%%%%%%%%%%%%%%%%%%%%%%%%%%%%%%%%%%%%%%%%%%%%%%%%%%%%%%%%%%%%%%%%%%%%%%%%%%%%%%%%%%%%%%%%%%%%%%%%%%%%%%%%%%%%%%%%%%%%%%%%%%%%%%%%%%%%%%%%%%%%%%%%%%%%%%%%%%%%%%%%%%%%%%%%%%%%%%%%%%%%%%%%%%%%%%%%%%%%%%%%%%%%%%%%%%%%%%%%%%%%%%%%%%%%%%%%%%%%%%%%%%%%%%%%%%%%%%%%%%%%%%%%%%%%%%%%%%%%%%%%%%%%%%%%%%%%

\section{Derivation of the Dyson expansion~\eqref{def:Dyson_setup}}

For the derivation, we first define
\begin{align}
U(\tau) := e^{\tau H} e^{-\tau (H + h)},
\end{align}
where we omit the index $i$ for the simplicity. 
By taking the differentiation of $U(\tau)$, we have
\begin{align}
\frac{d}{d\tau}U(\tau) &= H e^{\tau H} e^{-\tau (H + h)} - e^{\tau H} (H + h) e^{-\tau (H + h)}  \notag \\
&= \left[ H - e^{\tau H} (H + h) e^{-\tau H}\right]  e^{\tau H} e^{-\tau (H + h)}= - e^{\tau H} h e^{-\tau H}  U(\tau)
\end{align}
From $U(0)=\hat{1}$, we obtain
\begin{align}
U(\beta) = \mathcal{\tau} \left[e^{-\int_0^\beta e^{\tau H} h e^{-\tau H}d\tau}\right].
\end{align}

%%%%%%%%%%%%%%%%%%%%%%%%%%%%%%%%%%%%%%%%%%%%%%%%%%%%%%%%%%%%%%%%%%%%%%%%%%%%%%%%%%%%%%%%%%%%%%%%%%%%%%%%%%%%%%%%%%%%%%%%%%%%%%%%%%%%%%%%%%%%%%%%%%%%%%%%%%%%%%%%%%%%%%%%%%%%%%%%%%%%%%%%%%%%%%%%%%%%%%%%%%%%%%%%%%%%%%%%%%%%%%%%%%%%%%%%%%%%%%%%%%%%%%%%%%%%%%%%%%%%%%%%%%%%%%%%%%%%%%%%%%%%%%%%%%%%%%%%%%%%%%%%%%%%%%%%%%%%%%%%%%%%%%%%%%%%%%%%%%%%%%%%%%%%%%%%%%%%%%%%%%%%%%%%%%%%%%%%%%%%%%%%%%%%%%%%%%%%%%%%%%%%%%%%%%%%%%%%%%%%%%%%%%%%%%%%%%%%%%%%%%%%%%%%%%%%%%%%%%%%%%%%%%%%%%%%%%%%%%%%%%%%%%%%%%%%%%%%%%%%%%%%%%%%%%%%%%%%%%%%%%%%%%%%%%%%%%%%%%%%%%%%%%%%%%%%%%%%%%%%%%%%%%%%%%
%%%%%%%%%%%%%%%%%%%%%%%%%%%%%%%%%%%%%%%%%%%%%%%%%%%%%%%%%%%%%%%%%%%%%%%%%%%%%%%%%%%%%%%%%%%%%%%%%%%%%%%%%%%%%%%%%%

%Truncation formula

%%%%%%%%%%%%%%%%%%%%%%%%%%%%%%%%%%%%%%%%%%%%%%%%%%%%%%%%%%%%%%%%%%%%%%%%%%%%%%%%%%%%%%%%%%%%%%%%%%%%%%%%%%%%%%%%%%%%%%%%%%%%%%%%%%%%%%%%%%%%%%%%%%%%%%%%%%%%%%%%%%%%%%%%%%%%%%%%%%%%%%%%%%%%%%%%%%%%%%%%%%%%%%%%%%%%%%%%%%%%%%%%%%%%%%%%%%%%%%%%%%%%%%%%%%%%%%%%%%%%%%%%%%%%%%%%%%%%%%%%%%%%%%%%%%%%%%%%%%%%%%%%%%%%%%%%%%%%%%%%%%%%%%%%%%%%%%%%%%%%%%%%%%%%%%%%%%%%%%%%%%%%%%%%%%%%%%%%%%%%%%%%%%%%%%%%%%%%%%%%%%%%%%%%%%%%%%%%%%%%%%%%%%%%%%%%%%%%%%%%%%%%%%%%%%%%%%%%%%%%%%%%%%%%%%%%%%%%%%%%%%%%%%%%%%%%%%%%%%%%%%%%%%%%%%%%%%%%%%%%%%%%%%%%%%%%%%%%%%%%%%%%%%%%%%%%%%%%%%%%%%%%%%%%%%%%%%%%%%%%%%%%%%%%%%%%%%%%%%%%%%%%%%%%%%%%%%%%%%%%%%%%%%%%%%%%%%%%%%%%%%%%%%%%%%%%%%%%%%%%%%%%%%%%%%%%%%%%%%%%%%%%

\section{Derivation of the truncation formula~\eqref{approx_truncation_formulae}}

We here give the detail of the derivation of Eq.~\eqref{approx_truncation_formulae}.
From the definition $\rho(s):= e^{-\beta (H -s H_{\partial})}/\tr (e^{-\beta (H-s H_{\partial})})$, for $s=1$, we have $H -s H_{\partial}=H^{(L_l)}+H^{(L_l^\co)}$ and obtain
\begin{align}
\tr [ \rho(1) O_L] =\frac{\tr \left(e^{-\beta (H^{(L_l)}+H^{(L_l^\co)})}  O_L\right) }{\tr \bigl [ e^{-\beta (H^{(L_l)}+H^{(L_l^\co)})}\bigr]} 
= \frac{\tr \left(e^{-\beta H^{(L_l)}}  O_L\right) }{\tr \bigl [ e^{-\beta H^{(L_l)} }\bigr]} \cdot \frac{\tr \left(e^{-\beta H^{(L_l^\co)}}  \right) }{\tr \bigl [ e^{-\beta H^{(L_l^\co)} }\bigr]}   
= \tr ( \rho^{(L_l)} O_L) ,
\end{align}
where we use the fact that $O$ is defined on the site subset $L$.
Then, we need to calculate
 \begin{align}
\tr [\rho(1) O_L] - \tr (\rho O_L)&=\tr [ \rho(1) O_L] - \tr[ \rho(0) O_L]  =   \int_0^1 \frac{d}{ds}   \tr [ \rho(s) O_L]  ds \label{Eq0_truncate}
\end{align}
Now, the derivation of $\tr [ \rho(s) O] $ is given by
 \begin{align}
\frac{d}{ds}  \tr [ \rho(s) O_L]  &= \frac{1 }{\tr \bigl [ e^{-\beta (H -s H_{\partial})}\bigr]}  \frac{d}{ds} \tr \left(e^{-\beta (H -s H_{\partial})}  O_L\right)  
+  \tr \left(e^{-\beta (H -s H_{\partial})}  O_L\right)   \frac{d}{ds}  \frac{1 }{\tr \bigl [ e^{-\beta (H -s H_{\partial})}\bigr]}  \notag \\
&=\beta \int_0^1 \left\{ \tr [ \rho(s)^{\kappa} H_{\partial}\rho(s)^{1-\kappa} O_L ]   - \tr [ \rho(s)  O_L ] \tr [ \rho(s) H_{\partial}] \right\}  d\kappa   , \label{Eq1_truncate}
\end{align}
In the second equality, we use 
 \begin{align}
\frac{d}{ds} e^{-\beta (H -s H_{\partial})} = \int_0^1  e^{-\beta (H -s H_{\partial})(1-\kappa) }  ( \beta  H_{\partial} ) e^{-\beta (H -s H_{\partial})\kappa}   d\kappa. \label{Eq2_truncate}
\end{align}
By combining Eqs.~\eqref{Eq0_truncate}, \eqref{Eq1_truncate} and \eqref{Eq2_truncate}, we have
 \begin{align}
\tr ( \rho O_L)&=\tr ( \rho_1 O_L) -  \beta \int_0^1  \int_0^1 \left\{ \tr [\rho(s)^{1-\kappa}  H_{\partial} \rho(s)^{\kappa} O_L ]   - \tr [ \rho(s)  O_L ] \tr [ \rho(s) H_{\partial}] \right\}  d\kappa ds \notag \\
&=\tr ( \rho_1 O_L) -  \beta \int_0^1  \int_0^1 \Cor_{\rho(s)} (\rho(s)^{-\kappa} H_{\partial } \rho(s)^{\kappa},O_L)d\kappa ds. 
\end{align}

%%%%%%%%%%%%%%%%%%%%%%%%%%%%%%%%%%%%%%%%%%%%%%%%%%%%%%%%%%%%%%%%%%%%%%%%%%%%%%%%%%%%%%%%%%%%%%%%%%%%%%%%%%%%%%%%%%%%%%%%%%%%%%%%%%%%%%%%%%%%%%%%%%%%%%%%%%%%%%%%%%%%%%%%%%%%%%%%%%%%%%%%%%%%%%%%%%%%%%%%%%%%%%%%%%%%%%%%%%%%%%%%%%%%%%%%%%%%%%%%%%%%%%%%%%%%%%%%%%%%%%%%%%%%%%%%%%%%%%%%%%%%%%%%%%%%%%%%%%%%%%%%%%%%%%%%%%%%%%%%%%%%%%%%%%%%%%%%%%%%%%%%%%%%%%%%%%%%%%%%%%%%%%%%%%%%%%%%%%%%%%%%%%%%%%%%%%%%%%%%%%%%%%%%%%%%%%%%%%%%%%%%%%%%%%%%%%%%%%%%%%%%%%%%%%%%%%%%%%%%%%%%%%%%%%%%%%%%%%%%%%%%%%%%%%%%%%%%%%%%%%%%%%%%%%%%%%%%%%%%%%%%%%%%%%%%%%%%%%%%%%%%%%%%%%%%%%%%%%%%%%%%%%%%%%
%%%%%%%%%%%%%%%%%%%%%%%%%%%%%%%%%%%%%%%%%%%%%%%%%%%%%%%%%%%%%%%%%%%%%%%%%%%%%%%%%%%%%%%%%%%%%%%%%%%%%%%%%%%%%%%%%%

%Locality of the Belief-propagation 

%%%%%%%%%%%%%%%%%%%%%%%%%%%%%%%%%%%%%%%%%%%%%%%%%%%%%%%%%%%%%%%%%%%%%%%%%%%%%%%%%%%%%%%%%%%%%%%%%%%%%%%%%%%%%%%%%%%%%%%%%%%%%%%%%%%%%%%%%%%%%%%%%%%%%%%%%%%%%%%%%%%%%%%%%%%%%%%%%%%%%%%%%%%%%%%%%%%%%%%%%%%%%%%%%%%%%%%%%%%%%%%%%%%%%%%%%%%%%%%%%%%%%%%%%%%%%%%%%%%%%%%%%%%%%%%%%%%%%%%%%%%%%%%%%%%%%%%%%%%%%%%%%%%%%%%%%%%%%%%%%%%%%%%%%%%%%%%%%%%%%%%%%%%%%%%%%%%%%%%%%%%%%%%%%%%%%%%%%%%%%%%%%%%%%%%%%%%%%%%%%%%%%%%%%%%%%%%%%%%%%%%%%%%%%%%%%%%%%%%%%%%%%%%%%%%%%%%%%%%%%%%%%%%%%%%%%%%%%%%%%%%%%%%%%%%%%%%%%%%%%%%%%%%%%%%%%%%%%%%%%%%%%%%%%%%%%%%%%%%%%%%%%%%%%%%%%%%%%%%%%%%%%%%%%%%%%%%%%%%%%%%%%%%%%%%%%%%%%%%%%%%%%%%%%%%%%%%%%%%%%%%%%%%%%%%%%%%%%%%%%%%%%%%%%%%%%%%%%%%%%%%%%%%%%%%%%%%%%%%%%%%%

\section{Proof of the local approximation of $A_i$~\eqref{approx_A_i}}
In this section, we derive the inequality~\eqref{approx_A_i}.
We first restate it as follows: 
 \begin{align} 
\| A_i - A_i^{(L_l)}\|  \le C_{k,\beta}  \exp\left(\beta-\frac{\pi l}{2e k^3\beta}\right),  \label{approx_A_i_supp}
\end{align}
with $A_i=B_i B_i^\dagger$, $B_i = \mathcal{T}_\tau \bigl[e^{\int_0^1 \eta_i(\tau) d\tau } \bigr]$, and
 \begin{align} 
&\eta_i (\tau) = \frac{-\beta h_i}{2} - i \sum_{n\ge 1} \int_{-\infty}^{\infty} {\rm sign}(t) e^{-2\pi n t/\beta} h_i(t,\tau)dt \label{Eq.Belief_propagation}
\end{align}

By using the Suzuki-Trotter decomposition, the operator $B_i$ reduces to 
 \begin{align} 
B_i = \lim_{m\to \infty}  e^{\eta_i( \tau_m )/m} e^{ \eta_i( \tau_{m-1})/m}  \cdots  e^{ \eta_i( \tau_2)/m}e^{\eta_i(s \tau_1)/m}, \label{B_i_trotter_decomp}
\end{align}
where $\tau_s:=s/m$ for $s=1,2,\ldots,m$.
From the definition of $\eta_i(\tau)$, we have
 \begin{align} 
{\rm Re}\bigl[ \eta_i( \tau) \bigr] = \frac{-\beta h_i}{2} ,\quad {\rm Im}\bigl[ \eta_i( \tau) \bigr] = -  \sum_{n\ge 1} \int_{-\infty}^{\infty} {\rm sign}(t) e^{-2\pi n t/\beta} h_i(t,\tau)dt,
\end{align}
and hence, 
 \begin{align} 
\left \|e^{\eta_i( \tau_s )/m}\right\| &= \left \|e^{{\rm Re}\bigl[ \beta \eta_i( \tau_s )/m \bigr]} e^{i {\rm Im}\bigl[\beta \eta_i( \tau_s )/m \bigr]} \right\| + \orderof{m^{-2}}\notag \\
&  =  \left \|e^{- \beta h_i/(2m) \bigr]}\right\|+ \orderof{m^{-2}}  =   e^{\beta \|h_i\|/(2m)} + \orderof{m^{-2}}=e^{\beta/(2m)} + \orderof{m^{-2}}, \label{Norm_eta_i}
\end{align}
where we use $\|h_i\|=1$ in the last equality.
In the same way, we also obtain
 \begin{align} 
\left \|e^{\beta [\eta_i( \tau_s )]^{(L_l)} /m}\right\|  =e^{\beta/(2m)} + \orderof{m^{-2}}.\label{Norm_eta_i_L_l}
\end{align}
Note that ${\rm Re}\bigl\{ [\eta_i( \tau)]^{(L_l)}\bigr\} = -\beta h_i/2$.
By combining Eq.~\eqref{B_i_trotter_decomp} with Eqs.~\eqref{Norm_eta_i} and \eqref{Norm_eta_i_L_l}, we have 
 \begin{align} 
\| A_i - A_i^{(L_l)}\|  
\le &  e^{\beta} \lim_{m \to \infty} \sum_{s=1}^m \frac{1}{m} \left( \left \|  \eta_i( \tau_{s}) -  [\eta_i( \tau_{s})]^{(L_l)}   \right\|  + \left \|  \eta_i^\dagger( \tau_{s}) -  [\eta_i^\dagger( \tau_{s})]^{(L_l)}   \right\|   \right) \label{local_approx_A_i}
\end{align}

Now, we need to derive an upper bound for $\left \|  \eta_i( \tau_{s}) -  [\eta_i( \tau_{s})]^{(L_l)}   \right\|$.
For this purpose, we start from the Lieb-Robinson bound for $h_i(t,\tau)$. 
Let $O$ be an arbitrary operator which is separated from the operator $h_i$ by a distance $l$.
Then, the Lieb-Robinson bound reads 
  \begin{align} 
\left\|[ h_i(t,\tau), O] \right\|  \le  \frac{2|L|}{k} \cdot  \|h_i\| \cdot \|O\|   \frac{(2k |t|)^{m_0}}{m_0!} =2 \|O\|   \frac{(2k^2 |t|)^{m_0}}{m_0!}  \label{Lieb-Robinson_kuwahara}
\end{align}
with
  \begin{align} 
m_0:= \left \lfloor \frac{l}{k} +1 \right \rfloor \ge \frac{l}{k} 
\end{align}
Please see eq. (2.4) in Ref.~\cite{Phd_kuwa} for the derivation of the inequality~\eqref{Lieb-Robinson_kuwahara}.
Note that $h_i$ is supported in the subset $L:=\{i,i+1,\ldots,i+k-1\}$ with $|L|=k$ and the norm $\|h_i\|$ is equal to $1$.
By following the same discussion as in Ref.~\cite{PhysRevLett.97.050401}, we obtain
  \begin{align} 
 \left\|h_i(t,\tau) - [h_i(t,\tau)]^{(L_l)}\right\|  \le  C  \min \left( \frac{(2k^2 |t|)^{m_0}}{m_0!}, 1 \right) \le    C  \min \left[ \left(  \frac{2ek^2 |t|}{m_0} \right)^{m_0}, 1 \right] \label{local_approx_h_i_t_tau}
\end{align}
where $C$ is a constant of $\orderof{1}$, and in the second inequality, we use $m_0! \ge (m_0/e)^{m_0}$. 
Note that $[h_i(t,\tau)]^{(L_l)}$ is Hermitian from the definition~\eqref{truncation_in_a_region_from_l}.

By using the inequality~\eqref{local_approx_h_i_t_tau}, we have from Eq.~\eqref{Eq.Belief_propagation}
\begin{align} 
 \left\| \eta_i (\tau) - [ \eta_i (\tau)] ^{(L_l)}\right\| &\le 2 C \sum_{n\ge 1} \int_{0}^{\infty}e^{-2\pi n t/\beta} \min \left[   \left(  \frac{2ek^2 t}{m_0} \right)^{m_0}, 1 \right] dt \notag \\
 &= 2 C \int_{0}^{\infty} \frac{1}{e^{2\pi t/\beta}-1}\min \left[   \left(  \frac{2ek^2 t}{m_0} \right)^{m_0}, 1 \right] dt \label{local_approx_Eta_i}
\end{align}
When we define $t_0:= m_0/(2e k^2)$, $t_1:=\beta/(2\pi)$, the integral reduces to
\begin{align} 
&\int_{0}^{\infty} \frac{1}{e^{2\pi t/\beta}-1}\min \left[   \left(  \frac{2ek^2 t}{m_0} \right)^{m_0}, 1 \right] dt\le \int_{0}^{\infty} \left( 1+ \frac{t_1}{t} \right) e^{-t/t_1}  \min \left[  \left(  \frac{t}{t_0} \right)^{m_0} , 1 \right] dt\notag \\
&\le \int_{0}^{t_0/2}   \left( 1+ \frac{t_1}{t} \right)\left(  \frac{t}{t_0} \right)^{m_0} dt  + \int_{t_0/2}^\infty \left( 1+ \frac{2t_1}{t_0} \right) e^{-t/t_1}  dt \notag \\
&= 2^{-m_0} \left(\frac{t_0}{m_0+1} + \frac{2t_1}{m_0} \right) + t_1 \left( 1+ \frac{2t_1}{t_0} \right) e^{-t_0/(2t_1)} \le \tilde{C}_{k,\beta}\exp\left(-\frac{\pi l}{2e k^3\beta}\right) , \label{local_approx_Eta_i_integral}
\end{align}
where $\tilde{C}_{k,\beta}$ is a constant of $\orderof{k^2\beta^2}$.
From the inequalities~\eqref{local_approx_Eta_i} and \eqref{local_approx_Eta_i_integral}, we have
\begin{align} 
 \left\| \eta_i (\tau) - [ \eta_i (\tau)] ^{(L_l)}\right\| &\le 2C \tilde{C}_{k,\beta}\exp\left(-\frac{\pi l}{2e k^3\beta}\right) =:  \frac{C_{k,\beta}}{2}\exp\left(-\frac{\pi l}{2e k^3\beta}\right) \label{local_approx_Eta_i_final}
\end{align}
we have the same inequality for  $\left\| \eta_i^\dagger (\tau) - [ \eta_i^\dagger (\tau)] ^{(L_l)}\right\| $.
By applying the inequality~\eqref{local_approx_Eta_i_final} to \eqref{local_approx_A_i}, we obtain the inequality~\eqref{approx_A_i_supp}.

%%%%%%%%%%%%%%%%%%%%%%%%%%%%%%%%%%%%%%%%%%%%%%%%%%%%%%%%%%%%%%%%%%%%%%%%%%%%%%%%%%%%%%%%%%%%%%%%%%%%%%%%%%%%%%%%%%%%%%%%%%%%%%%%%%%%%%%%%%%%%%%%%%%%%%%%%%%%%%%%%%%%%%%%%%%%%%%%%%%%%%%%%%%%%%%%%%%%%%%%%%%%%%%%%%%%%%%%%%%%%%%%%%%%%%%%%%%%%%%%%%%%%%%%%%%%%%%%%%%%%%%%%%%%%%%%%%%%%%%%%%%%%%%%%%%%%%%%%%%%%%%%%%%%%%%%%%%%%%%%%%%%%%%%%%%%%%%%%%%%%%%%%%%%%%%%%%%%%%%%%%%%%%%%%%%%%%%%%%%%%%%%%%%%%%%%%%%%%%%%%%%%%%%%%%%%%%%%%%%%%%%%%%%%%%%%%%%%%%%%%%%%%%%%%%%%%%%%%%%%%%%%%%%%%%%%%%%%%%%%%%%%%%%%%%%%%%%%%%%%%%%%%%%%%%%%%%%%%%%%%%%%%%%%%%%%%%%%%%%%%%%%%%%%%%%%%%%%%%%%%%%%%%%%%%
%%%%%%%%%%%%%%%%%%%%%%%%%%%%%%%%%%%%%%%%%%%%%%%%%%%%%%%%%%%%%%%%%%%%%%%%%%%%%%%%%%%%%%%%%%%%%%%%%%%%%%%%%%%%%%%%%%

%bound of the norm of multicommutator

%%%%%%%%%%%%%%%%%%%%%%%%%%%%%%%%%%%%%%%%%%%%%%%%%%%%%%%%%%%%%%%%%%%%%%%%%%%%%%%%%%%%%%%%%%%%%%%%%%%%%%%%%%%%%%%%%%%%%%%%%%%%%%%%%%%%%%%%%%%%%%%%%%%%%%%%%%%%%%%%%%%%%%%%%%%%%%%%%%%%%%%%%%%%%%%%%%%%%%%%%%%%%%%%%%%%%%%%%%%%%%%%%%%%%%%%%%%%%%%%%%%%%%%%%%%%%%%%%%%%%%%%%%%%%%%%%%%%%%%%%%%%%%%%%%%%%%%%%%%%%%%%%%%%%%%%%%%%%%%%%%%%%%%%%%%%%%%%%%%%%%%%%%%%%%%%%%%%%%%%%%%%%%%%%%%%%%%%%%%%%%%%%%%%%%%%%%%%%%%%%%%%%%%%%%%%%%%%%%%%%%%%%%%%%%%%%%%%%%%%%%%%%%%%%%%%%%%%%%%%%%%%%%%%%%%%%%%%%%%%%%%%%%%%%%%%%%%%%%%%%%%%%%%%%%%%%%%%%%%%%%%%%%%%%%%%%%%%%%%%%%%%%%%%%%%%%%%%%%%%%%%%%%%%%%%%%%%%%%%%%%%%%%%%%%%%%%%%%%%%%%%%%%%%%%%%%%%%%%%%%%%%%%%%%%%%%%%%%%%%%%%%%%%%%%%%%%%%%%%%%%%%%%%%%%%%%%%%%%%%%%%%

\section{Proof of Lemma~\ref{Lem:Imaginary_lieb-Robinson}}

In the following, we derive Lemma~\ref{Lem:Imaginary_lieb-Robinson} on the imaginary Lieb-Robinson bound in one-dimensional systems.
For the purpose, we first derive the inequality~\eqref{ineq:norm_multi_commutator}, which gives the upper bound on the norm of multi-commutator.
We then prove Lemma~\ref{Lem:Imaginary_lieb-Robinson} by utilizing the bound.

\subsection{Derivation of an upper bound for the norm of multi-commutator}
We first derive an upper bound of the norm 
\begin{align}
\left\| \ad_{H^{(m)}}  \cdots \ad_{H^{(2)}} \ad_{H^{(1)}} (O_L) \right \| ,\label{1D_operators_ad_norm}
\end{align}
where $O_L$ is an operator defined on an adjacent subset $L$, and $\{H^{(s)}\}_{s=1}^m$ denote arbitrary one-dimensional operators as 
\begin{align}
H^{(s)}= \sum_{j=1}^n h^{(s)}_j \quad {\rm  with} \quad \|h^{(s)}_j \|\le 1 , \label{1D_operators_sup}
\end{align}
with each of $\{h_j\}_{j=1}^n$ acting on sites $\{j,j+1,\ldots,j+k-1\}$. 
In the following, we first derive Lemma~\ref{norm_multi_commutator_1D_1}, which bounds the norm~\eqref{1D_operators_ad_norm} in a general way. 
We then derive the inequality~\eqref{ineq:norm_multi_commutator} for the special case of Lemma~\ref{norm_multi_commutator_1D_1}.

Here, we prove the following Lemma:
\begin{lemma} \label{norm_multi_commutator_1D_1}
For arbitrary one-dimensional operators, $\{H^{(s)}\}_{s=1}^m$ given in Eq.~\eqref{1D_operators_sup}, the norm~\eqref{1D_operators_ad_norm} is bounded from above by 
\begin{align}
\left\| \ad_{H^{(m)}}  \cdots \ad_{H^{(2)}} \ad_{H^{(1)}} (O_L) \right \|  \le\|O_L\| \sum_{\pi_m=\pm,0}\cdots \sum_{\pi_2=\pm,0}\sum_{\pi_1=\pm,0} \prod_{s=1}^m 2 K_s(\pi_1,\pi_2,\ldots,\pi_m)  \label{basic_multi_com_1D_1}
\end{align}
with
\begin{align}
K_s(\pi_1,\pi_2,\ldots,\pi_m) =\begin{cases} 
\displaystyle {l_0+ \sum_{j\le s-1}(k-1) |\pi_j|  }& \for  \pi_s = 0 \\
k -1 & \for  \pi_s =\pm.
\end{cases} 
\label{Def_K_s_Leema_1D}
\end{align}
where we define $|\pi_j|=1$ if $\pi_j=\pm$.
We define $l_0$ as the number of sites contained in $L$ (i.e., $l_0=|L|$).
\end{lemma}

\begin{figure}[t]
\centering
{
\includegraphics[clip, scale=0.5]{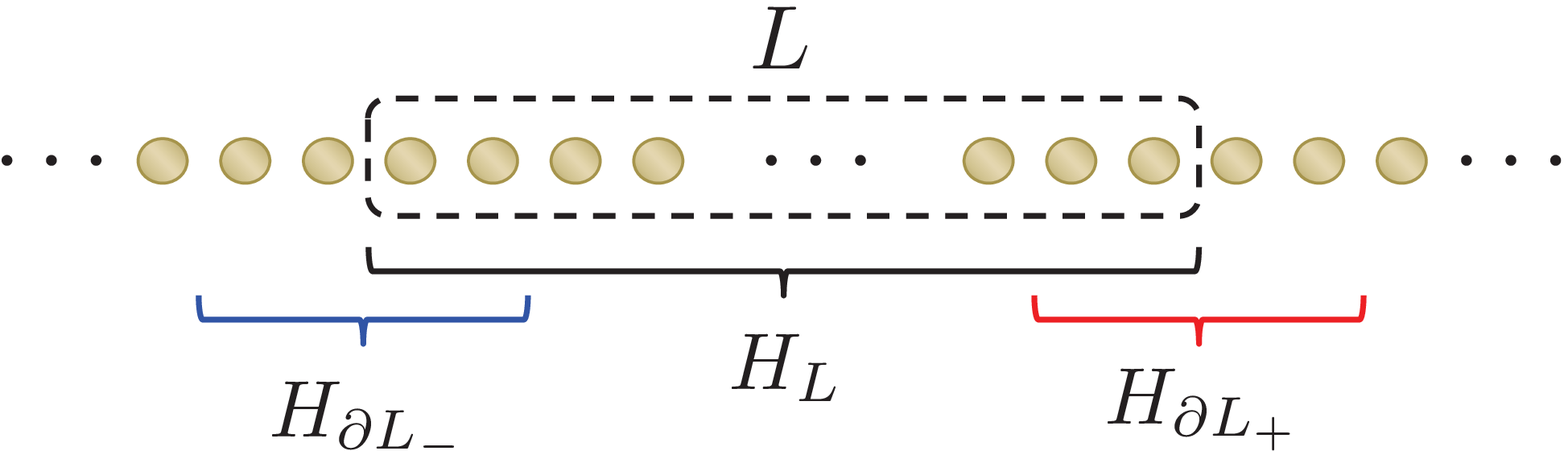}
}
\caption{}
\label{Fig_multi_commu}
\end{figure}

{~}\\
\textit{Proof of Lemma~\ref{norm_multi_commutator_1D_1}}.
%In the proof, we denote the boundary of subset $L$ as $\partial L$; for example, when $L=\{i_0, i_0+1,\ldots, i_0+l_0-1\}$, the boundary $\partial L$ is given by $\{i_0, i_0+l_0-1\}$.
We first decompose the Hamiltonian as follows
\begin{align}
H^{(1)}=H_L^{(1)} + H_{L^\co}^{(1)}+H_{\partial L_-}^{(1)}+H_{\partial L_+}^{(1)},
\end{align}
where $H_L^{(1)}$ and $H_{L^\co}^{(1)}$ act only the sites $L$ and $L^\co$, respectively, and $H_{\partial L_-}^{(1)}+H_{\partial L_+}^{(1)}$ denotes the interaction between the sites $L$ and $L^\co$.
The index $\pm$ discriminates the boundary interactions between the left side ($H_{\partial L_-}^{(1)}$) and the right side ($H_{\partial L_+}^{(1)}$) (Fig~\ref{Fig_multi_commu}).
We then obtain
\begin{align}
\ad_{H^{(1)}} (O_L) =\ad_{H_{\partial L_-}^{(1)}} (O_L) + \ad_{H_L^{(1)}} (O_L)+\ad_{H_{\partial L_+}^{(1)}} (O_L)
=:  O_{L_-}^{(1)} +O_{L_0}^{(1)} +O_{L_+}^{(1)}       ,         \label{decom_A_1_O_L}
\end{align}
where we use $\ad_{H_{L^\co}^{(1)}} (O_L)=0$. Note that $|L_0|=l_0$, $|L_-| \le l_0+ k-1$ and $|L_+| \le l_0+ k-1$.

We then estimate the norms of $O_{L_-}^{(1)}$, $O_{L_0}^{(1)}$ and $O_{L_+}^{(1)}$.
First, we obtain 
\begin{align}
\| O_{L_0}^{(1)}\| = \| \ad_{H_L^{(1)}} (O_L)\| 
\le \sum_{j\in L} \left\| \ad_{h_j^{(1)}} (O_L)\right\| 
\le 2\|O_L\|  \sum_{j\in L} \left\| h_j^{(1)} \right\|
 \le 2 |L|  \|O_L\| = 2l_0\|O_L\| ,\label{norm1_first_decomp_A_O_L}
\end{align}
where we used $\left\| h_j^{(1)} \right\| \le 1$ in the last inequality.
Second, because the interaction length is at most $k-1$, the number of $h_j$ which contributes to $H_{\partial L_-}^{(1)}$ is also at most $k-1$, and hence 
\begin{align}
\| O_{L_-}^{(1)}\| = \| \ad_{H_{\partial L_-}^{(1)}} (O_L)\| 
\le  2 (k-1) \|O_L\| .\label{norm1_first_decomp_A_O_L_minus}
\end{align}
The same inequality holds for $\| O_{L_+}^{(1)}\| $.

In the next step, we calculate $\ad_{H^{(2)}}\ad_{H^{(1)}} (O_L) $, which is now given by
\begin{align}
\ad_{H^{(2)}} \ad_{H^{(1)}} (O_L) =\ad_{H^{(2)}}\left(O_{L_-}^{(1)}\right) +\ad_{H^{(2)}}\left(O_{L_0}^{(1)}\right) +\ad_{H^{(2)}}\left(O_{L_+}^{(1)}\right)   )     ,         \label{decom_A_1_O_L2nd}
\end{align}
As in the calculation of $\ad_{H^{(1)}} (O_L) $, $\ad_{H^{(2)}}( O_{L_-}^{(1)})$ is decomposed into
\begin{align}
\ad_{H^{(2)}}( O_{L_-}^{(1)}) =\ad_{H_{\partial L_{--}}^{(2)}} \left(O_{L_-}^{(1)}\right) + \ad_{H_{L_-}^{(2)}}\left(O_{L_-}^{(1)}\right)  +\ad_{H_{\partial L_{-+}}^{(2)}} \left(O_{L_-}^{(1)}\right) 
=:  O_{L_{--}}^{(2)} +O_{L_{-0}}^{(2)} +O_{L_{-+}}^{(2)}       ,         
\end{align}
where we decompose $H^{(2)}= H_{L_-}^{(2)} + H_{L_-^\co}^{(2)}+H_{\partial L_{--}}^{(2)}+H_{\partial L_{-+}}^{(2)}$ with $H_{\partial L_{--}}^{(2)}$ and $H_{\partial L_{-+}}^{(2)}$ the boundary interactions of the left side and the right side, respectively.
The norms of $O_{L_{--}}^{(2)}$, $O_{L_{-0}}^{(2)}$ and $O_{L_{-+}}^{(2)}$ are also bounded from above by
\begin{align}
&\| O_{L_{--}}^{(2)} \|  \le  2 (k-1) \|O_{L_-}^{(1)}\| \le [2 (k-1)]^2 \|O_{L}\| , \quad \| O_{L_{-+}}^{(2)} \|  \le  2 (k-1) \|O_{L_-}^{(1)}\| \le [2 (k-1)]^2 \|O_{L}\| \notag \\
&\| O_{L_{-0}}^{(2)} \|  \le  2 |L_-| \|O_{L_-}^{(1)}\| \le 2 (l_0+k-1) \cdot 2(k-1)  \|O_{L}\| 
\end{align}
We apply the same calculations to $\ad_{H^{(2)}}(O_{L_0}^{(1)})$ and  $\ad_{H^{(2)}}(O_{L_+}^{(1)} )$, and obtain
\begin{align}
\ad_{H^{(2)}} \ad_{H^{(1)}} (O_L) = \sum_{\pi_1= \pm, 0} \sum_{\pi_2= \pm, 0}      O_{L_{\pi_1,\pi_2}}^{(2)}   
\end{align}
We notice that the norm of $O_{L_{\pi(1)\pi(2)}}^{(2)}$ satisfies  
\begin{align}
O_{L_{\pi_1,\pi_2}}^{(2)} \le K_1(\pi_1, \pi_2) K_2(\pi_1, \pi_2).
\end{align}

In the same way, we repeatedly decompose the multicommutator $\ad_{H^{(m)}}\cdots \ad_{H^{(2)}} \ad_{H^{(1)}} (O_L)$ into
\begin{align}
\ad_{H^{(m)}}\cdots \ad_{H^{(2)}} \ad_{H^{(1)}} (O_L) =\sum_{\pi_m= \pm, 0}\cdots \sum_{\pi_2= \pm, 0} \sum_{\pi_1= \pm, 0}  O_{L_{\pi_1,\pi_2,\ldots ,\pi_m}}^{(m)}   
\end{align}
and prove by induction that
\begin{align}
&\left | L_{\pi_1,\pi_2,\ldots \pi_m} \right | \le    l_0+ \sum_{j\le m}(k-1) |\pi_j| \notag \\
&\left\| O_{L_{\pi_1,\pi_2,\ldots ,\pi_m}}^{(m)} \right\| \le \prod_{s=1}^m 2 K_s(\pi_1,\pi_2,\ldots,\pi_m). \label{induction_multi_commu_1D}
\end{align}
For $m=1$, this is true from the above calculations.
Under the assumption that the inequalities~\eqref{induction_multi_commu_1D} hold for $m=m_0$, we need to prove the case of $m=m_0+1$.
By using the decomposition of the Hamiltonian~\eqref{decom_A_1_O_L}, we can obtain the same inequalities as \eqref{norm1_first_decomp_A_O_L} and \eqref{norm1_first_decomp_A_O_L_minus}  
for the commutator $\ad_{H^{(m_0+1)}} \left(O_{L_{\pi_1,\pi_2,\ldots ,\pi_{m_0}}}^{(m_0)}\right)$.
This yields the inequalities~\eqref{induction_multi_commu_1D} for the case of $m=m_0+1$.
This completes the proof. $\square$

From Lemma~\ref{norm_multi_commutator_1D_1}, we also obtain the following corollary:
\begin{corol} \label{norm_multi_commutator_1D_2}
The inequality~\eqref{basic_multi_com_1D_1} reduces to 
\begin{align}
\left\| \ad_{H^{(m)}}  \cdots \ad_{H^{(2)}} \ad_{H^{(1)}} (O_L) \right \|  \le\|O_L\| (6k)^m\left(\frac{m+\tilde{l}_0}{W[e(m+\tilde{l}_0)]}\right)^{m+\tilde{l}_0-\frac{m+\tilde{l}_0}{W[e(m+\tilde{l}_0)]}} 
\label{ineq:norm_multi_commutator_1D_2}
\end{align}
with
$
\tilde{l}_0:= l_0/k.
$
\end{corol}
Particularly, for $l_0\le k$ or $\tilde{l}_0\le 1$, we have 
\begin{align}
\left(\frac{m+\tilde{l}_0}{W[e(m+\tilde{l}_0)]}\right)^{m+\tilde{l}_0-\frac{m+\tilde{l}_0}{W[e(m+\tilde{l}_0)]}} \le  \left(\frac{m}{\gamma \log m}\right)^m  \label{special_case_multi_com}
\end{align}
with $\gamma$ a constant of $\gamma \simeq 1.6026$. 
By combining the inequalities~\eqref{ineq:norm_multi_commutator_1D_2} and \eqref{special_case_multi_com}, we obtain the inequality \eqref{ineq:norm_multi_commutator}.

{~}\\
\textit{Proof of Corollary~\ref{norm_multi_commutator_1D_1}}.
In the proof, we need to estimate the upper bound of
\begin{align}
\sum_{\pi_m=\pm,0}\cdots \sum_{\pi_2=\pm,0}\sum_{\pi_1=\pm,0} \prod_{s=1}^m 2 K_s(\pi_1,\pi_2,\ldots,\pi_m)  .
\end{align}
For the purpose, we define $N_{m,q}$ with $q$ a integer ($0\le q\le m$) as 
\begin{align}
N_{m,q}:= \sum_{|\pi_1|+ |\pi_2| + \cdots + |\pi_m|=q } \prod_{s=1}^m 2 K_s(\pi_1,\pi_2,\ldots,\pi_m) .
\end{align}
Note that we defined $|\pi_s|=1$ if $\pi_s=\pm$.
For a fixed $q$, we have $l_0+\sum_{j\le s-1}(k-1)| \pi_j|  \le l_0 + q(k-1)$, and hence
\begin{align}
N_{m,q}&\le  \sum_{\pi_m(1)+ \pi_m(2) + \cdots + \pi_m(m)=q }2^m \left[ l_0 + q(k-1)\right]^{m-q}  (k-1)^{q} \notag \\
&=  \binom{m}{q}2^{m+q} \left[ l_0 + q(k-1)\right]^{m-q}  (k-1)^{q} \le  \binom{m}{q}2^{m+q} k^m (q+\tilde{l}_0)^{m-q}  
\end{align}
where in the second equality we use $\sum_{|\pi_1|+ |\pi_2| + \cdots + |\pi_m|=q }=2^{q}\binom{m}{q} $, and  in the last inequality we define $\tilde{l}_0:=l_0/k$.
Then, we obtain
\begin{align}
\sum_{\pi_m} \prod_{s=1}^m 2 K_s = \sum_{q=0}^m N_{m,q}\le  (2k)^m \sum_{q=0}^m   \binom{m}{q} 2^q (q+\tilde{l}_0)^{m-q} . \label{Ineq_W_func_before}
\end{align}
For $(q+\tilde{l}_0)^{m-q}$, we can derive the following inequality by solving a standard maximization problem:
\begin{align}
(q+\tilde{l}_0)^{m-q}  \le \left(\frac{m+\tilde{l}_0}{W[e(m+\tilde{l}_0)]}\right)^{m+\tilde{l}_0-\frac{m+\tilde{l}_0}{W[e(m+\tilde{l}_0)]}}, 
\end{align}
where $W(x)$ is the Lambert W function, which is defined by $W(x)e^{W(x)}=x$. 
This reduces the inequality~\eqref{Ineq_W_func_before} to
\begin{align}
\sum_{\pi_m} \prod_{s=1}^m 2 K_s \le  (6k)^m\left(\frac{m+\tilde{l}_0}{W[e(m+\tilde{l}_0)]}\right)^{m+\tilde{l}_0-\frac{m+\tilde{l}_0}{W[e(m+\tilde{l}_0)]}} 
\end{align}
This completes the proof. $\square$

%%%%%%%%%%%%%%%%%%%%%%%%%%%%%%%%%%%%%%%%%%%%%%%%%%%%%%%%%%%%%%%%%%%%%%%%%%%%%%%%%%%%%%%%%%%%%%%%%%%%%%%%%%%%%%%%%%%%%%%%%%%%%%%%%%%%%%%%%%%%%%%%%%%%%%%%%%%%%%%%%%%%%%%%%%%%%%%%%%%%%%%%%%%%%%%%%%%%%%%%%%%%%%%%%%%%%%%%%%%%%%%%%%%%%%%%%%%%%%%%%%%%%%%%%%%%%%%%%%%%%%%%%%%%%%%%%%%%%%%%%%%%%%%%%%%%%%%%%%%%%%%%%%%%%%%%%%%%%%%%%%%%%%%%%%%%%%%%%%%%%%%%%%%%%%%%%%%%%%%%%%%%%%%%%%%%%%%%%%%%%%%%%%%%%%%%%%%%%%%%%%%%%%%%%%%%%%%%%%%%%%%%%%%%%%%%%%%%%%%%%%%%%%%%%%%%%%%%%%%%%%%%%%%%%%%%%%%%%%%%%%%%%%%%%%%%%%%%%%%%%%%%%%%%%%%%%%%%%%%%%%%%%%%%%%%%%%%%%%%%%%%%%%%%%%%%%%%%%%%%%%%%%%%%%%
%%%%%%%%%%%%%%%%%%%%%%%%%%%%%%%%%%%%%%%%%%%%%%%%%%%%%%%%%%%%%%%%%%%%%%%%%%%%%%%%%%%%%%%%%%%%%%%%%%%%%%%%%%%%%%%%%%

%Imaginary Lieb-Robinson bound

%%%%%%%%%%%%%%%%%%%%%%%%%%%%%%%%%%%%%%%%%%%%%%%%%%%%%%%%%%%%%%%%%%%%%%%%%%%%%%%%%%%%%%%%%%%%%%%%%%%%%%%%%%%%%%%%%%%%%%%%%%%%%%%%%%%%%%%%%%%%%%%%%%%%%%%%%%%%%%%%%%%%%%%%%%%%%%%%%%%%%%%%%%%%%%%%%%%%%%%%%%%%%%%%%%%%%%%%%%%%%%%%%%%%%%%%%%%%%%%%%%%%%%%%%%%%%%%%%%%%%%%%%%%%%%%%%%%%%%%%%%%%%%%%%%%%%%%%%%%%%%%%%%%%%%%%%%%%%%%%%%%%%%%%%%%%%%%%%%%%%%%%%%%%%%%%%%%%%%%%%%%%%%%%%%%%%%%%%%%%%%%%%%%%%%%%%%%%%%%%%%%%%%%%%%%%%%%%%%%%%%%%%%%%%%%%%%%%%%%%%%%%%%%%%%%%%%%%%%%%%%%%%%%%%%%%%%%%%%%%%%%%%%%%%%%%%%%%%%%%%%%%%%%%%%%%%%%%%%%%%%%%%%%%%%%%%%%%%%%%%%%%%%%%%%%%%%%%%%%%%%%%%%%%%%%%%%%%%%%%%%%%%%%%%%%%%%%%%%%%%%%%%%%%%%%%%%%%%%%%%%%%%%%%%%%%%%%%%%%%%%%%%%%%%%%%%%%%%%%%%%%%%%%%%%%%%%%%%%%%%%%%

\subsection{Derivation of the imaginary Lieb-Robinson bound}

In this section, we prove the imaginary Lieb-Robinson bound which is given as Lemma~\ref{Lem:Imaginary_lieb-Robinson}.
We restate the Lemma as follows:
\begin{lemma} \label{Lem:Imaginary_lieb-Robinson_re}
For an arbitrary one-dimensional Hamiltonian $H$ in the form of Eq.~\eqref{eq:1D_ham}, we have
 \begin{align} 
 \|O_L(i\tau) - [O_L(i\tau)]^{(L_l)} \|  \le \frac{\zeta_l ^{\lceil l/k \rceil}}{1-\zeta_l},  \label{approx_LR}
\end{align}
with $O_L(i\tau) := e^{\tau H}O_L e^{-\tau H}$ and
\begin{align}
\zeta_l := \frac{6e k\tau}{\gamma \log \lceil l/k \rceil }, \label{def_of_xi_L}
\end{align}
where $O_L$ is an arbitrary operator supported in an adjacent subset $L$ ($|L|\le k$).
\end{lemma}

\textit{Proof of Lemma~\ref{norm_multi_commutator_1D_1}.}
For the proof, we apply the Baker-Campbell-Hausdorff expansion as 
\begin{align}
O_L(i\tau)= e^{\tau H}O_L e^{-\tau H}= \sum_{m=0}^{\infty}\frac{\tau^{m}}{m!} \ad_{H}^{m} (O_L)
\end{align}
Because the interactions in $H$ act at most $k$ adjacent sites, any terms $\ad_{H}^{m} (O_L)$ with $m \le l/k$ is supported in the subset $L_l$.
Hence, we need to estimate 
\begin{align}
\left\| O_L(i\tau)-O_L^{(L_l)}(\beta_1, \beta_2) \right\| \le  \sum_{m> l/k}^{\infty}\frac{\tau^{m}}{m!}\left\| \ad_{H}^{m} (O_L)\right\|
\end{align}
By using Corollary~\ref{norm_multi_commutator_1D_1} with the inequality~\eqref{special_case_multi_com}, we have 
\begin{align}
\left\| \ad_{H}^{m} (O_L)\right\| \le \|O_L\| \left(\frac{6k m}{\gamma \log m}\right)^{m} \le  m! \left(\frac{6e k}{\gamma \log m}\right)^{m},
\end{align}
where we use the inequality $m! \ge (m/e)^m$.
We thus obtain
\begin{align}
\sum_{m> l/k}^{\infty}\frac{\tau^{m}}{m!}\left\| \ad_{H}^{m} (O_L)\right\|
\le& \sum_{m> l/k}^{\infty} \left(\frac{6e k \tau}{\gamma \log m}\right)^{m}\le \frac{\zeta_l ^{\lceil l/k \rceil}}{1-\zeta_l}, \label{summation_with_respect_to_m1_m2}
\end{align}
where $\zeta_l$ is defined in Eq.~\eqref{def_of_xi_L}.
This completes the proof. $\square$

\end{widetext}

\end{document}